\documentclass[preprint,nofootinbib]{revtex4-1}
\usepackage[paper=letterpaper,margin=1.5in]{geometry}

\usepackage{epsfig}
\usepackage{graphics}\graphicspath{{graph/}{}}
\usepackage{amssymb,amsmath,amsfonts}
\usepackage{time}
\usepackage{epstopdf}
\usepackage{color}
\usepackage{hyperref}
\hypersetup{
  colorlinks,
  citecolor=magenta,
  linkcolor=blue}

\newcommand{\p}{\partial}

\newcommand{\parder}[2]{\frac{\partial #1}{\partial #2}}

\renewcommand{\Re}{{\rm Re\,}}

\renewcommand{\vec}[1]{\textnormal{\boldmath$#1$}}

\begin{document}

\thispagestyle{empty}
\renewcommand{\thefootnote}{\fnsymbol{footnote}}

\begin{flushright}
{\small
SLAC-PUB-16171 \\
December 2014\\}
\end{flushright}

\vspace{.8cm}

\begin{center}
{\bf\large Using pipe with corrugated walls for a sub-terahertz FEL}

\vspace{1cm}

Gennady Stupakov\\
SLAC National Accelerator Laboratory,\\
2575 Sand Hill Road, Menlo Park, CA 64025

\medskip

\end{center}
\vfill

\begin{center}
{\it Submitted for publication to Physical Review Special Topics - Accelerators and Beams}
\end{center}

\newpage

%
\section{Introduction}
%

For applications in fields as diverse as chemical  and  biological imaging, material science,       telecommunication, semiconductor and superconductor research, there is great interest in having a source of intense pulses of terahertz radiation. Laser-based sources of such radiation~\cite{Auston:84,You:93} are capable of generating several-cycle pulses with frequency over the range 10--70~THz and energy of 20~$\mu$J~\cite{Sell:08}. In a beam-based sources, utilizing short, relativistic electron bunches~\cite{Nakazato:89,Carr:02}  an electron bunch impinges on a thin metallic foil and generates coherent transition radiation (CTR). An implementation of this method at the Linac Coherent Light Source (LCLS) has obtained single-cycle pulses of radiation that is broad-band, centered on 10~THz, and contains $>0.1$~mJ of energy~\cite{Daranciang:11}. Another beam-based method generates THz radiation by passing a bunch through a metallic pipe with a dielectric layer. As reported in~\cite{Cook:09}, this method was used to generate narrow-band pulses with frequency 0.4~THz and energy 10~$\mu$J. 

It has been noted in the past, in the study of wall-roughness impedance~\cite{bane99n,bane00st}, that a metallic pipe with corrugated walls supports propagation of a high-frequency mode that is in resonance with a relativistic beam. This mode can be excited by a beam whose length is a fraction of the wavelength. Similar to the dielectric-layer method, metallic pipe with corrugated walls can serve as a source of terahertz radiation~\cite{terahertz12}.

In this paper we study another option of excitation of the resonant mode in a metallic pipe with corrugated walls---via the mechanism of the free electron laser instability. This mechanism works if the bunch length is much longer than the wavelength of the radiation. While our focus will be on a metallic pipe with corrugated walls, most our results are also applicable to a dielectric-layer round geometries. The connection between the electrodynamic properties of the two types of structures can be found in Ref.~\cite{stupakov12b}.

Our analysis is carried out for relativistic electron beams with the Lorentz factor $\gamma \gg 1$. However, in some places we will keep small terms on the order of $1/\gamma^2$ to make our results valid for relatively moderate values of $\gamma \sim 5-10$. In particular, we will take into account that the particles' velocity $v$ differs from the speed of light $c$ (in contrast to the approximation $v=c$ typically made in~\cite{bane99n,bane00st,terahertz12}) . We will see that the FEL mechanism becomes much less efficient in the limit $\gamma \to \infty$, so the moderate values of $\gamma$ are of particular interest.

This paper is organized as follows. In Section~\ref{sec:2} we discuss the resonant frequency, the group velocity and the loss factor of the resonant mode whose phase velocity is equal to the velocity of the particle. Their derivations are given in Appendices~\ref{app:1} and~\ref{app:2}. In section~\ref{sec:3} we find the gain length and an estimate for the saturated power of an FEL in which a relativistic beam excites the resonant mode. In section~\ref{sec:4} we consider a practical numerical example of such an FEL. In section~\ref{sec:5} we discuss some of the effects that are not included in our analysis.

%
\section{Wake in a round pipe with corrugated walls}\label{sec:2}
%

We consider a round metallic pipe with inner radius $a$. 
\begin{figure}[htb]
\centering
\includegraphics[height=0.4\textwidth, trim=0mm 0mm 0mm 0mm, clip]{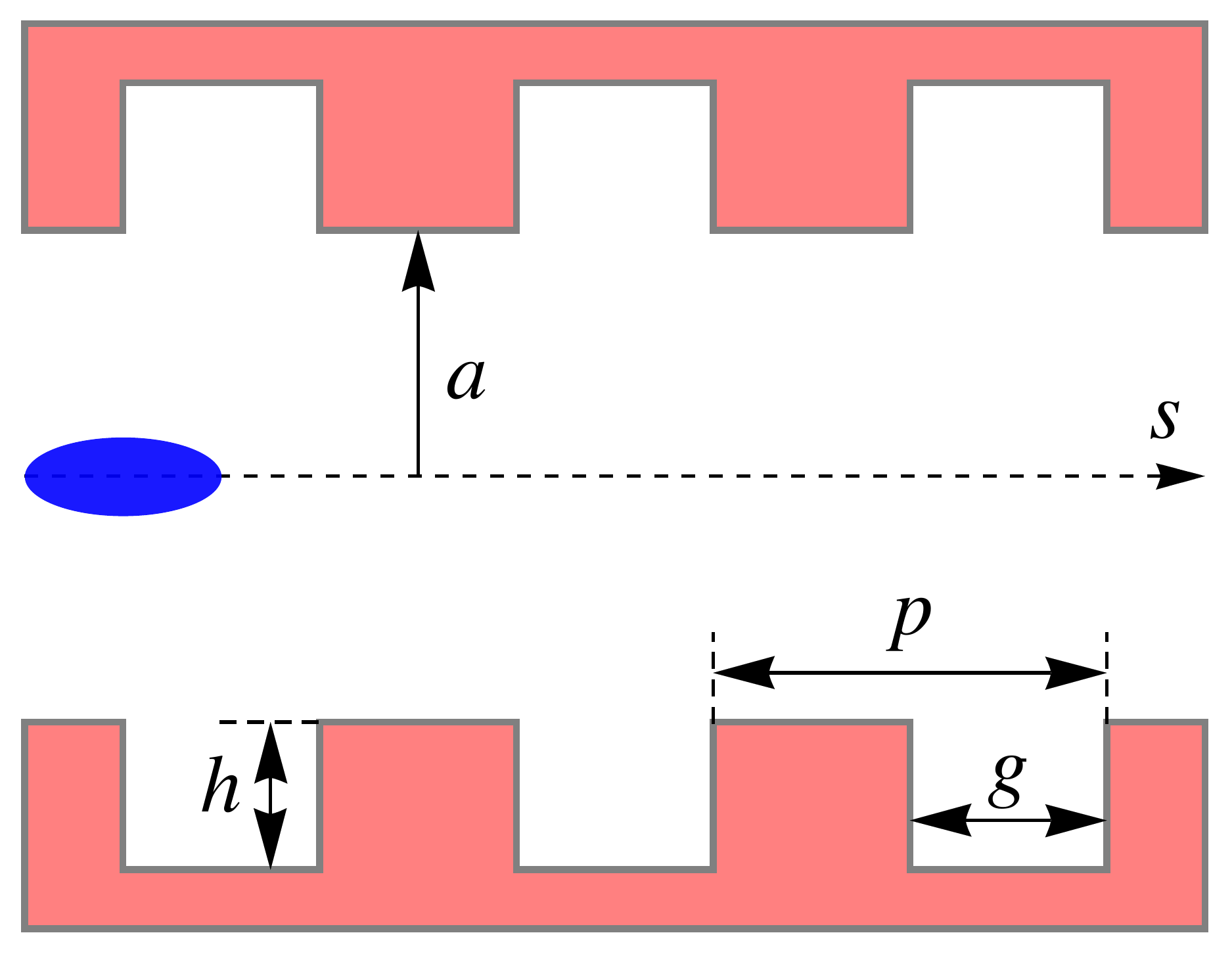}
\hspace{2mm}
\includegraphics[height=0.4\textwidth, trim=0mm 0mm 0mm 0mm, clip]{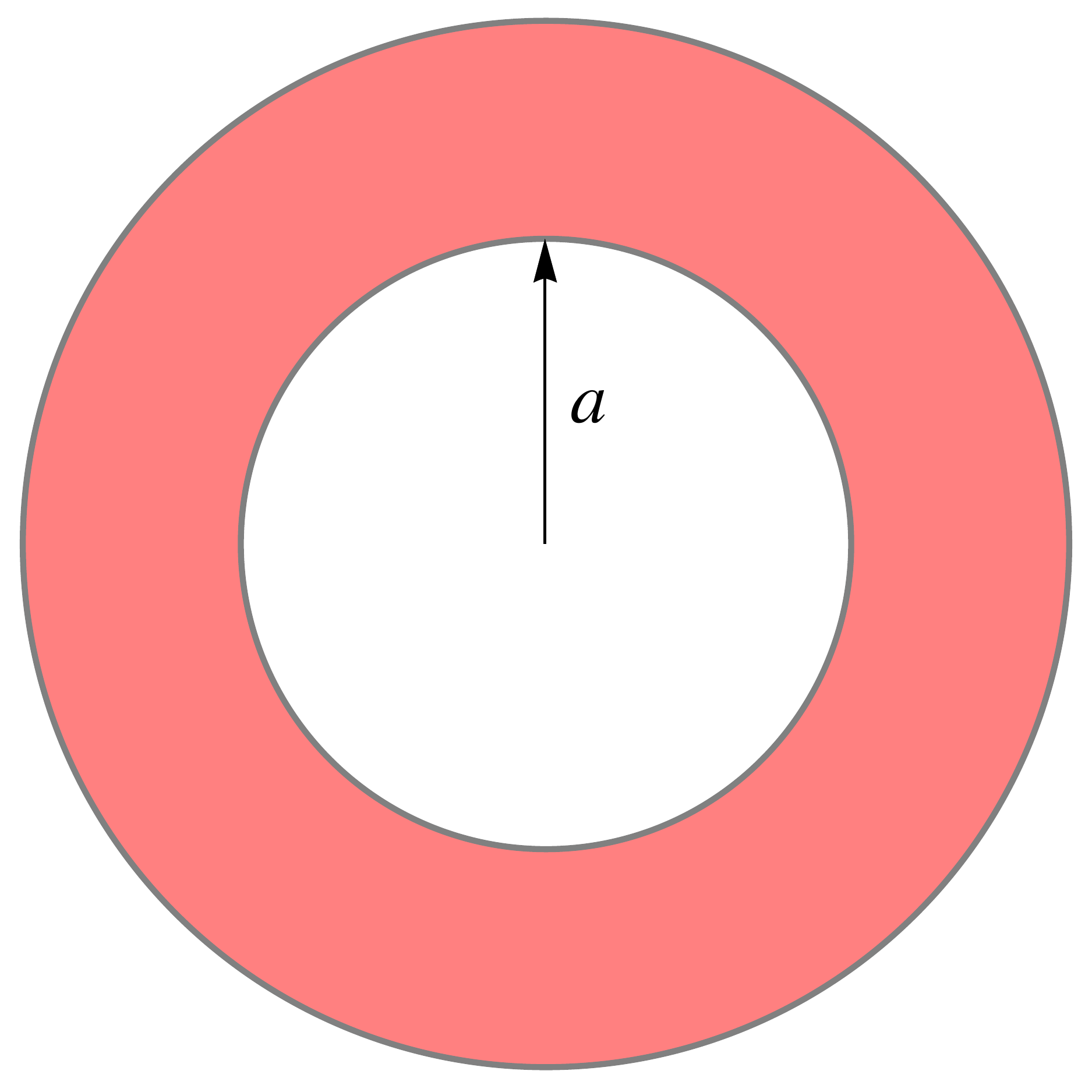}
\caption{Dimensions of a round corrugated pipe. An electron beam propagates along the axis of the pipe. The beam position $s$ in the pipe is measured along the axis with $s=0$ corresponding to the entrance to the pipe.}
\label{fig:1}
\end{figure}
Small rectangular corrugations have depth $h$, period $p$ and gap $g$, as shown in Fig.~\ref{fig:1}. In the case when $h,p\ll a$ and $h\gtrsim p$, the fundamental resonant mode with the phase velocity equal to the speed of light, $v_{ph}=c$, has the frequency $\omega_0 =ck_0$ and the group velocity $v_{g0}$, where~\cite{bane99n,bane00st}
    \begin{align}\label{eq:1}
    k_0
    =
    \left(
    \frac{2p}{agh}
    \right)^{1/2}
    ,\qquad
    1
    -
    \frac{v_{g0}}{c}
    =
    \frac{4gh}{ap}
    .
    \end{align}
Such a mode will be excited by an ultra-relativistic particle moving along the axes of the pipe with velocity $v=c$. Note that from the assumption $h,p\ll a$ follows the high-frequency nature of the resonant mode, $k_0\gg 1/a$.

\begin{figure}[htb]
\centering
\includegraphics[width=0.6\textwidth, trim=0mm 0mm 0mm 0mm, clip]{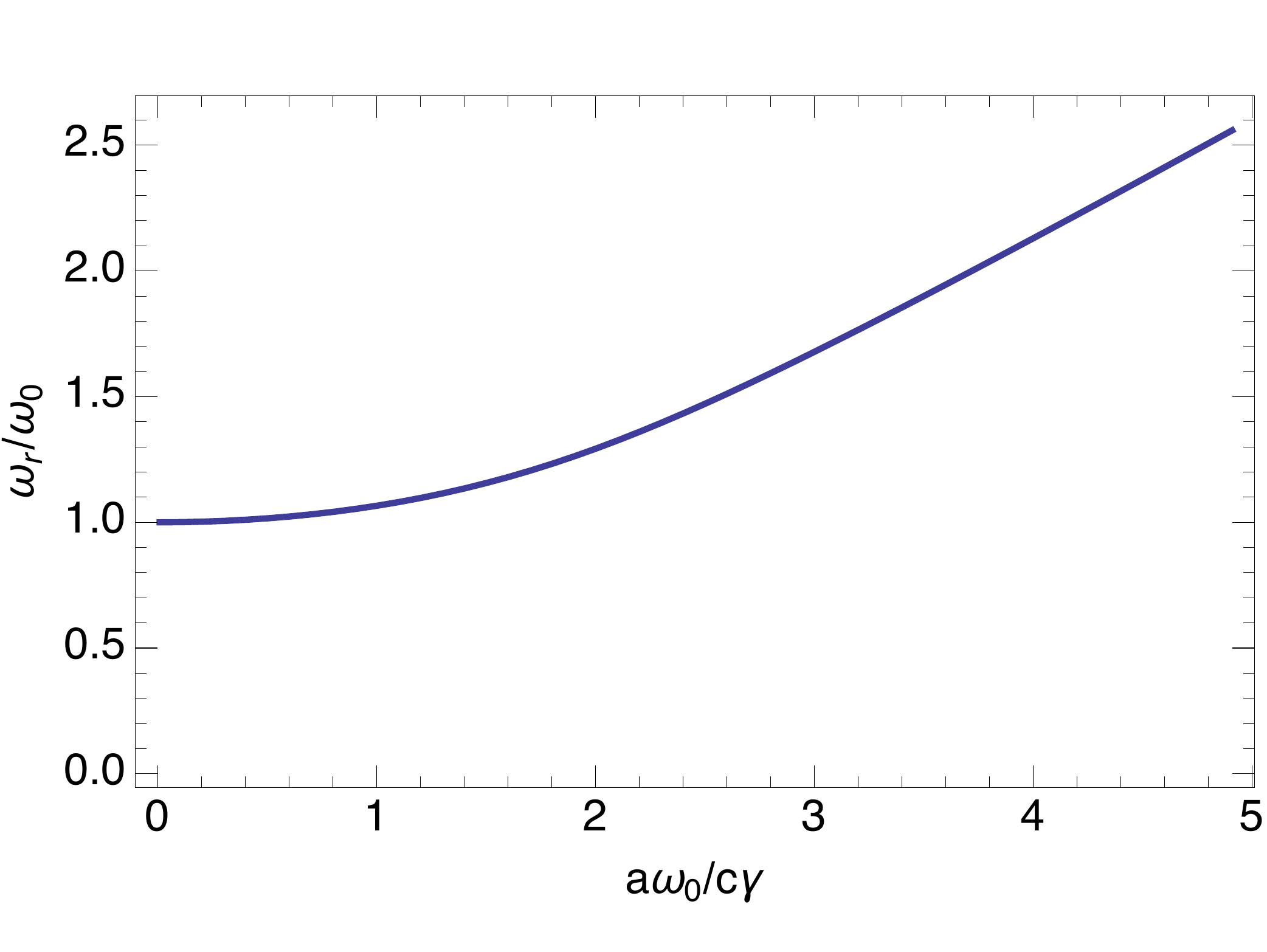}
\caption{Plot of the normalized frequency $\omega_r$ of the resonant wave as a function of the parameter $a\omega_0/c\gamma$.}
\label{fig:2}
\end{figure}

\begin{figure}[htb]
\centering
\includegraphics[width=0.6\textwidth, trim=0mm 0mm 0mm 0mm, clip]{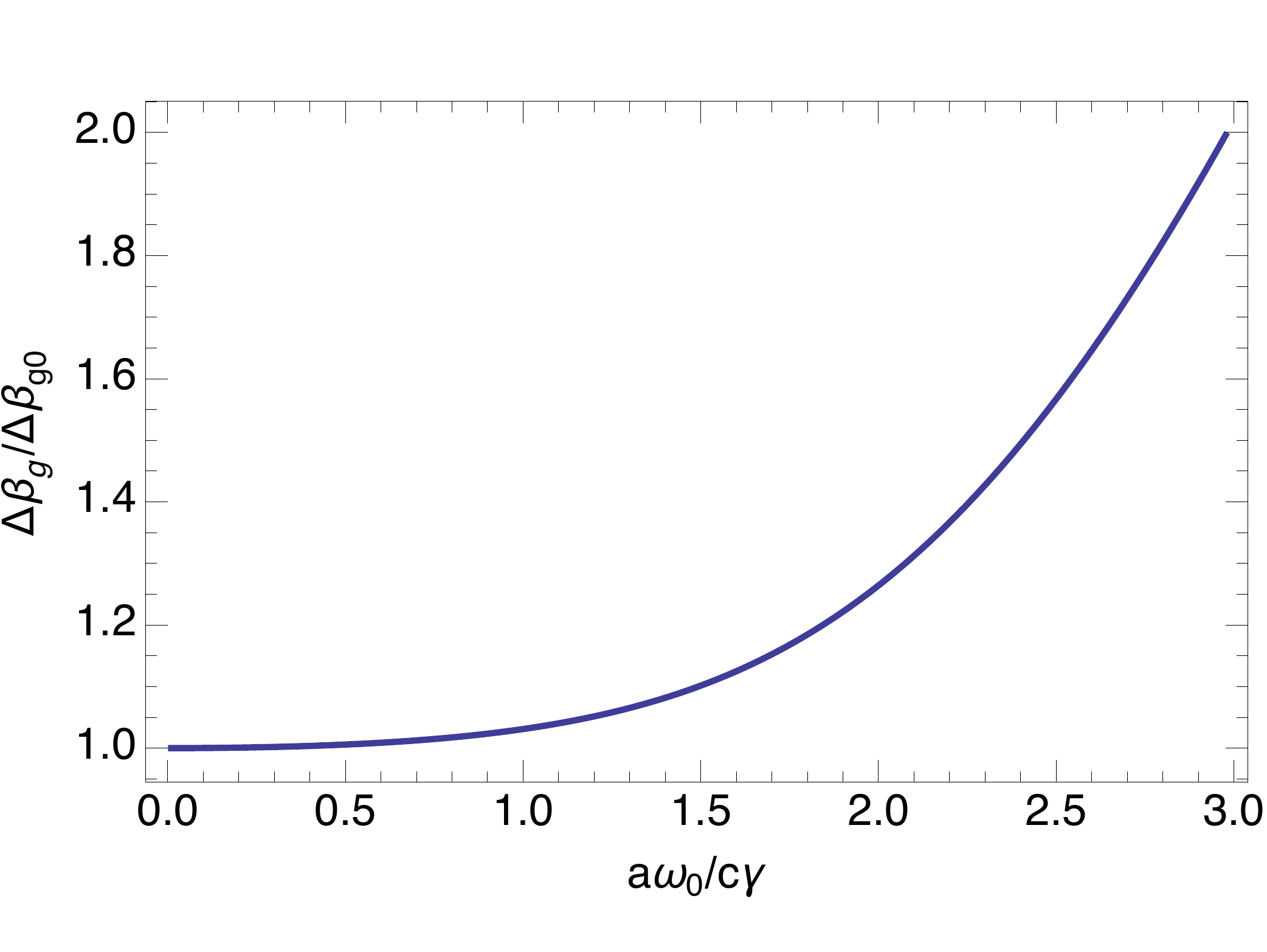}
\caption{Plot of the ratio  $\Delta\beta_g/\Delta\beta_{g0}$ (with $\Delta\beta_{g0} = 1-v_{g0}/c$ defined in~\eqref{eq:1}) versus the parameter $a\omega_0/c\gamma$. }
\label{fig:3}
\end{figure}

As explained in the Introduction, in our analysis we would like to take into account the fact that the phase velocity of the resonant mode is smaller than the speed of light, $v_{ph} = v<c$. Calculation of the frequency and the group velocity of the resonant mode for this case is carried out in Appendix~\ref{app:1}. As follows from this calculation, the deviation of the resonant frequency and the group velocity from Eqs.~\eqref{eq:1} is controlled by the parameter
    \begin{align}\label{eq:2}
    u = \frac{ak_0}{\gamma}
    \end{align}
with $k_0$ defined by~\eqref{eq:1}. The plot of the frequency $\omega_r$ of the resonant mode versus parameter $u$  is shown in Fig.~\ref{fig:2}.
We see that decreasing the beam energy $\gamma$ increases the frequency $\omega_r$ of the mode. Note  that because $k_0a\gg 1$ the deviation from the ultra-relativistic results~\eqref{eq:1} can become important even for large values of gamma, $\gamma\sim k_0a$. The group velocity of the resonant mode for $u\sim 1$ also deviates from the limit $\gamma\to\infty$ given by~\eqref{eq:1}. Calculations of the group velocity are given in Appendix~\ref{app:1} and the plot of $\Delta\beta_g = 1-v_g/c$ versus $u$ is shown in Fig.~\ref{fig:3}.

A relativistic point charge entering the pipe at the longitudinal coordinate $s=0$ and moving along the pipe axis excites the resonant mode and generates a longitudinal wakefield. The standard description of this process in accelerator physics is based on the notion of the (longitudinal) wake $w(z)$ that depends on the distance between the source and the test charges measured in the direction of motion \cite{chao93}. In case of the resonant mode, this wake is localized behind the driving charge and is equal to $w(z) = 2\varkappa \cos(\omega_r z/c)$ where $\varkappa$ is the loss factor per unit length (see, e.g.,~\cite{stupakov12b,stupakov_bane_dechirper12}). For our purposes, it is important to modify this wake taking into account that at any given distance $s$ from the entrance to the pipe, the wake extends behind the particle over a finite length; this makes the wake a function of two variables, $w(s,z)$. The distance at which the wake extends behind the charge can be obtained from a simple consideration: the wake  propagates with the group velocity $v_g$ and when the charge travels distance $s$ with speed $v$ the wake emitted at $s=0$ lags behind the charge at the distance $\Delta z = s(1-v_g/v)$ (we assume $v_g<v$).  Mathematically, this is expressed by the following equation:
    \begin{align}\label{eq:3}
    w(s,z)
    =
    \left\{
    \begin{array}
    {rl}
    2\varkappa \cos(\omega_r z/c),& \mathrm{ for }\,\, -s(1-v_g/v) < z < 0\\
    \varkappa,& \mathrm{ for }\,\,  z = 0\\
    0,& \mathrm{otherwize}
    \end{array}
    \right.
    .
    \end{align}
The sign of the wake~\eqref{eq:3} is such that a positive wake corresponds to the energy loss, and a negative wake means the energy gain. Note that the wake is only non-zero for negative $z$, that is behind the source charge. 

The loss factor $\varkappa_0$ in the limit $\gamma\to\infty$ is given by~\cite{stupakov_bane_dechirper12}
    \begin{align}\label{eq:4}
    \varkappa_0
    =
    \frac{2}{a^2}
    .
    \end{align} 
With account of finite, but large, value of $\gamma$ the loss factor is derived in Appendix~\ref{app:2}. It is plotted in Fig.~\ref{fig:4} again as a function of parameter $u$.
\begin{figure}[htb]
\centering
\includegraphics[width=0.6\textwidth, trim=0mm 0mm 0mm 0mm, clip]{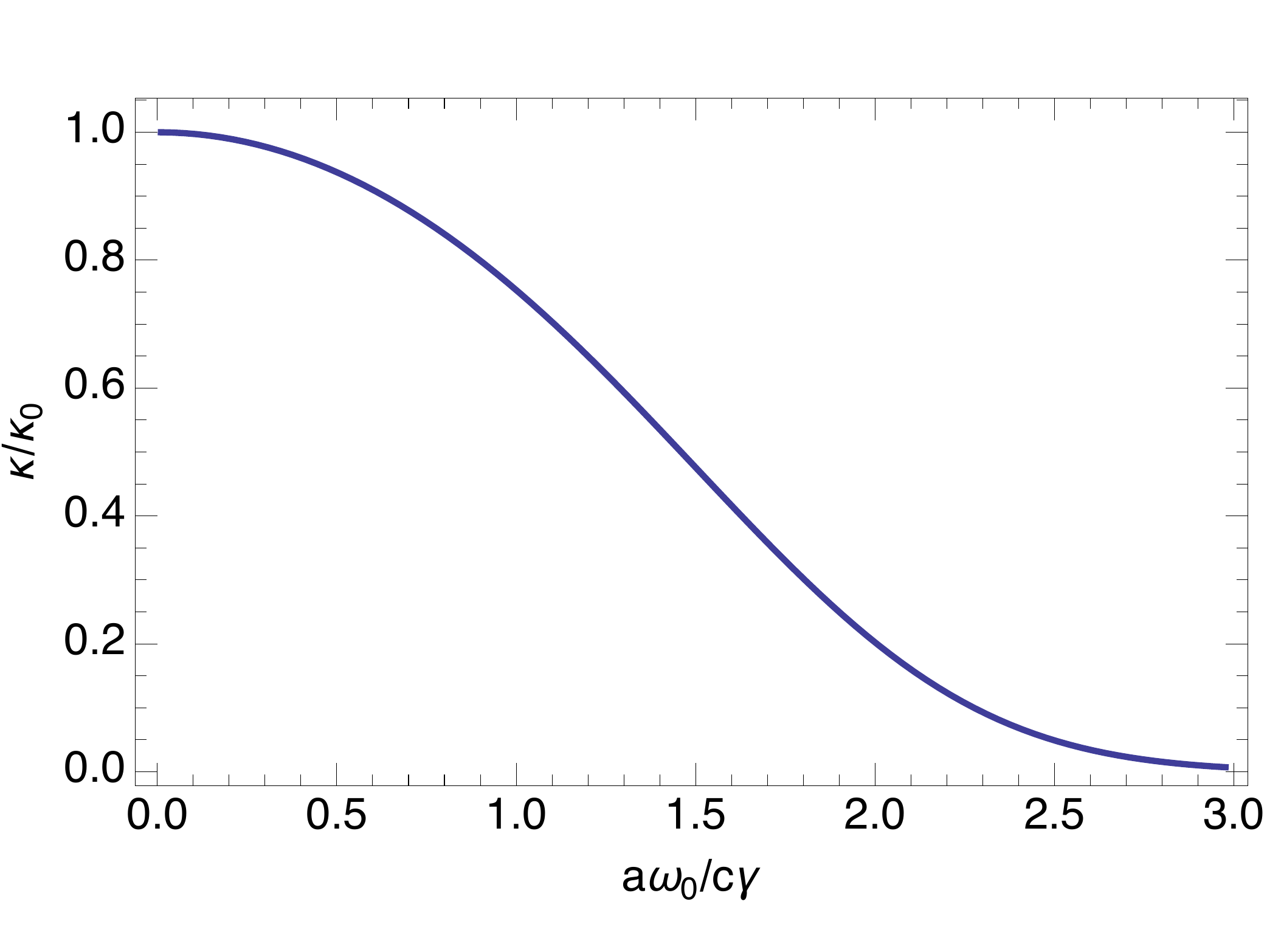}
\caption{Plot of the normalized loss factor $\varkappa/\varkappa_0$ factor versus parameter $u = ak_0/\gamma$.}
\label{fig:4}
\end{figure}
We see that the interaction of the mode with the beam decreases when $\gamma$ becomes small. This happens because the spot size of the relativistically compressed Coulomb field of the point charge field on the wall of the pipe has the size on the order of $a/\gamma$, and when $u\sim 1$, is comparable with the inverse wave number of the wake $c/\omega_0$. For $u\gtrsim 1$ the frequency content of the Coulomb field at wavenumbers $\sim\omega_0/c$ gets depleted, and the excitation of the resonant mode is suppressed.

%
\section{1D FEL equations}\label{sec:3}
%

We now consider an electron beam of energy $\gamma mc^2$ with the transverse size much smaller than the pipe radius $a$ and with the uniform longitudinal current distribution propagating along a pipe with corrugated walls. Such a beam will be driving a resonant mode in the pipe, and if the pipe is long enough, it will become modulated and micro-bunched through the interaction with the mode. The mechanism of this interaction is exactly the same as in the free electron laser instability. In this section we describe an approach to calculate this instability, following the method developed in  Ref.~\cite{PAC03stupakov03kr}. The actual derivation is presented in Appendix~\ref{app:3}.

The crucial step in the derivation is a modification of the standard Vlasov equation that describes evolution of the distribution function of the beam. This modification takes into account retardation effects associated with emission of the wake field. The distribution function of the beam $f(\eta,z,s)$ is a function of the relative energy deviation, $\eta = \Delta \gamma/\gamma_0$, with $\gamma_0$ corresponding to the averaged beam energy, longitudinal position inside the bunch $z$, and the distance $s$ from the entrance to the pipe. The evolution of $f$ is described by the Vlasov equation\
    \begin{align}\label{eq:5}
    &
    \frac{\partial f}{\partial s}
    -
    \alpha\eta 
    \frac{\partial f}{\partial z}
    -
    \frac{r_0}{\gamma}
    \frac{\partial f}{\partial \eta}
    \int_{-\infty}^\infty dz'
    \int_{-\infty}^\infty d\eta'
    w(s,z-z')
    f\left(
    \eta',z',s-v\frac{z'-z}{v-v_g}
    \right)
    =
    0
    ,
    \end{align}
where $\alpha = -\gamma^{-2}$ is the slip factor per unit length and $r_0 = e^2/mc^2$ is the classical electron radius. The distribution function $f$ is normalized so that $\int fd\eta$ gives the number of particles per unit length.  The third argument of $f$ in the integrand of~\eqref{eq:5} takes into account the retardation: the wake that is generated by a beam slice at coordinate $z'$ slips behind the slice with the velocity $v-v_g$ relative to the beam, and if it reaches the point $z$ when the beam arrives at location $s$, it should have been emitted at position $s-v(z'-z)/(v-v_g)$ ~\cite{PAC03stupakov03kr}.

To establish a closer analogy with the standard FEL theory, it is convenient to  introduce a new variable $k_w$ (an analog of the FEL undulator wave number) defined by the equation
    \begin{align}\label{eq:6}
    \frac{k_0}{k_w}
    =
    \frac{v}{v-v_g}
    \approx
    \frac{1}{\Delta \beta_g - \Delta \beta_{ph}}
    ,
    \end{align}
where $\Delta \beta_g = 1 - v_g/c$ and $\Delta \beta_{ph} = 1 - v_{ph}/c$. In the ultra-relativistic limit $\gamma\to\infty$ using~\eqref{eq:1} we find
    \begin{align}\label{eq:7}
    k_w
    =
    k_{w0}
    \equiv
    4
    \left(
    \frac{2gh}{a^3p}
    \right)^{1/2}
    .
    \end{align}

Eq.~\eqref{eq:5} is linearized assuming a small perturbation of the beam equilibrium $f_0(\eta)$, $f=f_0(\eta) + f_1(\eta, z, s)$, with $|f_1| \ll f_0$. In this analysis we assume a coasting beam with the equilibrium distribution function $f_0(\eta) = n_0 F(\eta)$, where $n_0$ is the number of particles per unit length of the beam. We seek the perturbation in the form $f_1\propto e^{ikz+q k_w s}$, where $k$ is the wavenumber and $q$ is the dimensionless propagation constant whose real part is responsible for the exponential growth (or decay, if $\Re q<0$) of the perturbation with $s$. The main result of the linear instability analysis is the dispersion relation that defines the propagation constant $q$ as a function of the frequency detuning $\nu = (ck - \omega_r)/\omega_r$. This dispersion relation is derived in Appendix~\ref{app:3} (it follows closely the derivation of Ref.~\cite{PAC03stupakov03kr}), and is given by~\eqref{eq:53},
    \begin{align}\label{eq:8}
    \frac{1}{2}
    \frac{(2\rho)^3}{q - i\nu}
    \int_{-\infty}^\infty 
    d\eta
    \frac{F'(\eta)}{q - i\alpha\eta({\omega_r}/{ck_w})}
    =
    1
    \,,
    \end{align}
where the parameter $\rho$ (an analog of the Pierce parameter \cite{bonifacio84pn}) is
    \begin{align}\label{eq:9}
    (2\rho)^3
    = 
    \frac{2n_0\kappa c r_0}{k_w\gamma\omega_r}
    .
    \end{align}
Except for a slight notational difference, Eqs.~\eqref{eq:8} and~\eqref{eq:9} coincide with the standard equations of the 1D FEL theory~\cite{huang07k}.

For a cold beam, $F(\eta) = \delta(\eta)$ (here $\delta$ stands for the delta-function), and from~\eqref{eq:8} we obtain
    \begin{align}\label{eq:10}
    q^2(q - i\nu)
    =
    -
    \frac{i
    \alpha\omega_r}{2ck_w}
    (2\rho)^3
    .
    \end{align}
If follows from this equation that the fastest growth of the instability is achieved at zero detuning. Assuming $\nu=0$ we rewrite~\eqref{eq:10} using the definition~\eqref{eq:9} and $\alpha = -1/\gamma^2$,
    \begin{align}\label{eq:11}
    q^3
    =
    i
    \frac{ n_0\kappa r_0}{k_w^2\gamma^3}
    .
    \end{align}
Among the three roots of this equation, there is one, which we denote $q_1$, with a positive real part. Introducing the power gain length $\ell = (2\Re q_1 k_w)^{-1}$, and using $n_0r_0 = I/I_A$, where $I$ is the beam current and $I_A = 17.5$ kA is the Alfven current, we obtain
    \begin{align}\label{eq:12}
    \ell
    =
    \frac{1}{\sqrt{3}}
    \gamma
    \left(
    {\kappa k_w}
    \frac{I}{I_A}
    \right)^{-1/3}
    .
    \end{align}

In addition to the gain length, an important characteristic of the described FEL is the radiation power at saturation. Here we can use the result of the standard FEL theory, that the saturation occurs at the distance equal about 10-20 gain length, and the saturation power $P_{\mathrm{sat}}$ is
    \begin{align}\label{eq:13}
    P_{\mathrm{sat}}
    \approx
    \rho\gamma mc^2
    \frac{I}{e}
    .
    \end{align}
In the next section we will consider a practical example of an FEL based on a pipe with corrugated walls and evaluate $\ell$ and $P_{\mathrm{sat}}$ for that example.

%
\section{Numerical example}\label{sec:4}
%

To give an illustrative example of a practical device we consider in this section a pipe with corrugated walls with the parameters close to those accepted in Ref.~\cite{terahertz12}. Noting from Eq.~\eqref{eq:12} that the gain length is proportional to the beam energy, and having in mind a compact device, we choose a relatively small beam energy of 5 MeV. The beam current is 100 A. The pipe and corrugation dimensions with the beam parameters are summarized in Table~\ref{tab:1}. 
\begin{table}[hbt]
   \centering
   \caption{Corrugation and beam parameters}
   \begin{tabular}{|l|c|}\hline\hline
         Pipe radius, mm         & 2\\
         Depth $h$, $\mu$m       & 50\\
         Period $p$, $\mu$m      & 40\\
         Gap $g$, $\mu$m         & 10\\
         Bunch charge, nC        & 1\\
         Energy, MeV             & 5\\
         Bunch length, ps        & 10\\
         \hline\hline
   \end{tabular}
   \label{tab:1}
\end{table}
Note that parameter $u$ defined by~\eqref{eq:2} is $u=1.3$, and hence the deviation from the ultra-relativistic limit (corresponding to $u\ll 1$) is expected to be noticeable.

From Eq.~\eqref{eq:26} we find that the frequency $\omega_r/2\pi$ of the resonant mode is $0.34$ THz. Using the results of the Appendices~\ref{app:1} and~\ref{app:2} we find the group velocity of the resonant mode, $\Delta \beta_g = 0.053$, and the loss factor $\kappa = 0.6(2/a^2)= 2.7$ kV/(pC m), and calculate the Pierce parameter $\rho=0.013$. This gives the gain length $\ell \approx 7$ cm, and the saturation power $P_{\mathrm{sat}} \approx 6.7$ MW.

It is interesting to point out that for a given pipe radius and corrugations, there is an optimal value of the beam energy that minimized the gain length. This follows from Eq.~\eqref{eq:12} which shows that $\ell$ increases with $\gamma$ due to an explicit dependence $\ell\propto \gamma$, but $\ell$ also increases when $\gamma$ becomes too small due to the decrease of $\kappa$ shown in Fig.~\ref{fig:4}. As numerical minimization shows, the minimal value or $\ell$ is achieved for $u=1.9$ and is given by
    \begin{align}\label{eq:15}
    \ell
    =
    0.74
    \frac{a^2k_0}{2\sqrt{3}}
    \left(
    \frac{I_A}{I}
    \right)^{1/3}
    \left(
    \frac{ap}{2hg}
    \right)^{1/6}
    .
    \end{align}
For the parameter considered above this gives the optimal value of the beam energy: $\gamma = 6.6$ with the corresponding gain length $\ell = 5.5$ cm.

%
\section{Discussion}\label{sec:5}
%

There are several issues of practical importance that were omitted in our analysis in preceding sections. Here will briefly discuss some of them leaving a more detailed study for a separate publication.

First, we used an approximation of a coasting beam, without taking into account the finite length of the bunch. This approximation assumes that the bunch length is much longer than the cooperation length of the instability $l_\mathrm{c}$ that is defined as the distance at which the point charge wake extends within the bunch when  the particle travels one gain length $\ell$. Using Eq.~\eqref{eq:3} we evaluate the coherence length as $l_\mathrm{c}\sim \ell(1-v_g/v)$. For the parameters considered in Section~\ref{sec:4} we find $l_\mathrm{c}\approx 3.3$ mm, or 11 ps. This is comparable with the bunch length of 10 ps, and hence the numerical estimates of the previous section should only be considered as crude estimates of the expected parameters of the FEL.  A more accurate prediction for the selected set of parameters require computer simulations. 

Second, we neglected the resistive wall losses that would cause the resonant mode to decay when it propagates in the pipe. The effect of the wall losses on the FEL instability can be estimated if we compare the gain length with the decay distance $l_\mathrm{d}$ of the resonant mode. An analytical formula for $l_\mathrm{d}$ is given in Ref.~\cite{terahertz12}; using the formula we estimate that for our parameters  $l_\mathrm{d}=66$ cm, which is much larger than the gain length calculated in the previous section. Hence, we conclude that the resistive wall effect is small.

Finally, we mention a deleterious effect of the transverse wake, that might cause the beam break-up instability. It is known that in a round pipe with corrugated walls, in addition to the resonant longitudinal wake, there is also a resonant dipole mode that creates a transverse wakefield. In the limit $\gamma\to\infty$, in a round pipe, the transverse mode has the same frequency as the longitudinal one. To mitigate the effect of the breakup instability, one has to apply a strong external transverse focusing on the beam and minimize the initial beam offset at the entrance to the pipe. It may also be advantageous to change the cross sections of the pipe from round to rectangular or elliptic, that will likely detune the transverse mode frequency from the longitudinal one. A more detailed study of the transverse instability is necessary.

%
\section{Acknowledgments}
%

The author thanks M. Zolotorev and K. Bane and I. Kotelnikov for useful discussions.

This work was supported by Department of Energy contract DE-AC03-76SF00515.

\appendix
%
\section{Resonant mode for moderate values of $\gamma$}\label{app:1}
%

In this Appendix we analyze properties of the resonant mode in a round pipe with corrugated walls assuming $\gamma\gg 1$ but keeping small terms on the order of $\gamma^{-2}$. The resonant mode in this case is defined as a mode that has the phase velocity $v_{ph} = c\sqrt{1-\gamma^{-2}}$. Our analysis is performed for the steady state wakefield; the modification due to the finite interaction length is done straightforwardly using Eq.~\eqref{eq:3}. 

It is shown in Ref.~\cite{stupakov12b} that small wall corrugations can be treated as a thin material layer with  some effective values of the dielectric permeability $\epsilon$ and magnetic permittivity $\mu$. Calculations of $\epsilon$ and $\mu$ for given values of the corrugation parameters are carried out in~\cite{stupakov12b} where it is shown that $\mu = g/p$ and the effective dielectric permeability $\epsilon$ is typically small and can be neglected in comparison with $\mu$.  The electrodynamical properties of the layer are expressed through the surface impedance $\zeta$ that relates the longitudinal component of the electric field with the azimuthal magnetic field on the wall,
    \begin{align}\label{eq:16}
    E_z|_{r=a}
    =
    -
    \zeta
    H_\theta|_{r=a}
    ,
    \end{align}
where~\cite{stupakov12b}
    \begin{align}\label{eq:17}
    \zeta(\omega,k_z)
    =
    ih
    \frac{\omega}{c}
    \left(
    \frac{k_z^2c^2}{\omega^2}\epsilon^{-1}
    -
    \mu
    \right)
    .
    \end{align}

To find the resonant mode we write an axisymmetric TM-like solution of Maxwell's equations in the pipe with the time and $z$ dependences $\propto e^{-i\omega t +ik_z z}$ in the following form
    \begin{align}\label{eq:18}
    E_z
    =
    E_0
    I_0(k_r r)
    ,\qquad
    H_\theta
    =
    -
    E_0
    \frac{i\omega}{ck_r}
    I_1(k_r r)
    ,
    \end{align}
where $E_0$ is the field amplitude and
    \begin{align}\label{eq:19}
    k_r
    =
    \sqrt{k_z^2-\frac{\omega^2}{c^2}}
    =
    \frac{\omega}{c}
    (\beta_{ph}^{-2}-1)^{1/2}
    .
    \end{align}
Here $\beta_{ph} = v_{ph}/c$ with $v_{ph} = \omega/k_z$ the phase velocity of the wave, $I_0$ and $I_1$ are the modified Bessel functions of the first kind, and we assume $k_z>\omega/c$ so that  $\beta_{ph} < c$, and $k_r$ is real.

We now substitute~\eqref{eq:18} into the boundary condition~\eqref{eq:16}, \eqref{eq:17} to obtain
    \begin{align}\label{eq:20}
    k_r a
    \frac{
    I_0(k_r a)
    }
    {
    I_1(k_r a)
    }
    =
    \zeta
    \frac{ia\omega }{c}
    .
    \end{align}
Taking into account that the phase velocity is close to the speed of light, $1-\beta_{ph} \ll 1$, we will use for $\zeta$ a simplified equation in which $k_z^2c^2/\omega^2$ us replaced by unity,
    \begin{align}\label{eq:21}
    \zeta\approx {ih\omega}{c^{-1}}(\epsilon^{-1}-\mu)
    .
    \end{align}
From~\eqref{eq:20} we find
    \begin{align}\label{eq:22}
    x
    \frac{
    I_0(x)
    }
    {
    I_1(x)
    }
    =
    \frac{ha\omega^2 }{c^2}
    (\mu-\epsilon^{-1})
    ,
    \end{align}
with 
    \begin{align}\label{eq:23}
    x
    =
    k_r a
    =
    a\sqrt{k_z^2-\frac{\omega^2}{c^2}}
    .
    \end{align}

Consider first an ultra-relativistic limit $\beta_{ph} \to 1$. In this limit, $x\to 0$ and     $    \lim_{x\to 0}    x    {    I_0(x)    }/    {    I_1(x)    }    =    2  $. Substituting this into~\eqref{eq:22} we recover the standard result for the synchronous mode
    \begin{align}\label{eq:24}
    \frac{\omega_r}{c}
    =
    \frac{\omega_0}{c}
    \equiv
    \left[
    \frac{2}{ha(\mu-\epsilon^{-1})}
    \right]^{1/2}
    .
    \end{align}
Eq.~\eqref{eq:1} is obtained from this expression by substituting $\mu = g/p$ and neglecting $\epsilon$ (see details in~\cite{stupakov12b}).

We now assume $x\sim 1$ and write it as
    \begin{align}\label{eq:25}
    x
    =
    a
    \frac{\omega}{c}
    (\beta_{ph}^{-2}-1)^{1/2}
    \approx
    \frac{a\omega_r}{c \gamma}
    ,
    \end{align}
where we used the resonant mode condition $\beta_{ph} = \sqrt{1-\gamma^{-2}} \approx 1-\frac{1}{2}\gamma^{-2}$. Using the notations $u={a\omega_0}/{c \gamma}$ and $y_r=\omega_r/\omega_0$ we rewrite~\eqref{eq:22},
    \begin{align}\label{eq:26}
    u
    \frac{I_0(uy_r)}{I_1(uy_r)}
    =
    2y_r
    .
    \end{align}
This equation was solved numerically and the dependence $y_r(u)$ is plotted in Fig.~\ref{fig:2}.

When $x\sim 1$ the group velocity of the resonant wave also deviates from the value given by the second equation in~\eqref{eq:1}. To find the group velocity we first differentiate~\eqref{eq:22} with respect to $\omega$:
    \begin{align}\label{eq:27}
    \frac{dx}{d\omega}
    \frac{d}{dx}
    x
    \frac{
    I_0(x)
    }
    {
    I_1(x)
    }
    =
    \frac{2ha\omega }{c^2}
    (\mu-\epsilon^{-1})
    ,
    \end{align}
and then use~\eqref{eq:23} to find $dx/d\omega$,
    \begin{align}\label{eq:28}
    \frac{dx}{d\omega}    
    =    
    \frac{a^2\omega}{xc^2}
    \left(
    \frac{1}{\beta_{ph}\beta_g}
    -
    1
    \right)  
    ,
    \end{align}
where $\beta_g = v_g/c = c^{-1}d\omega/dk_z$. Combining~\eqref{eq:27} and~\eqref{eq:28} yields
    \begin{align}\label{eq:29}
    \left(
    \frac{1}{\beta_{ph}\beta_g}
    -
    1
    \right)
    =
    \frac{2h }{a}
    (\mu-\epsilon^{-1})
    \left(
    \frac{1}{x}
    \frac{d}{dx}
    x
    \frac{I_0(x)}
    {I_1(x)}
    \right)^{-1}
    .
    \end{align}
We now use $\beta_{ph}\approx 1-\frac{1}{2}\gamma^{-2}$ and $\beta_g = 1-\Delta\beta_g$ and recalling that $\gamma \gg 1$ and $\Delta\beta_g \ll 1$ obtain
    \begin{align}\label{eq:30}
    \Delta\beta_g
    =
    \frac{2h }{a}
    (\mu-\epsilon^{-1})
    \left(
    \frac{1}{x}
    \frac{d}{dx}
    x
    \frac{I_0(x)}
    {I_1(x)}
    \right)^{-1}
    \bigg|_{x=uy_r}
    -
    \frac{1}{2\gamma^2}
    .
    \end{align}
In the limit $\gamma\to \infty$ we have $x\to 0$ and one can find from~\eqref{eq:30}
    \begin{align}\label{eq:31}
    \lim_{\gamma\to\infty}\Delta\beta_g
    =
    \Delta\beta_{g0}
    \equiv
    \frac{4h}{a}
    (\mu-\epsilon^{-1}).
    \end{align}
Again, neglecting $\epsilon$ and substituting $\mu=g/p$ one recovers the group velocity in Eq.~\eqref{eq:1}. In the general case, we normalize $\Delta\beta_g$ by $\Delta\beta_{g0}$,
    \begin{align}\label{eq:32}
    \frac{\Delta\beta_g}{\Delta\beta_{g0}}
    =
    \frac{1}{2}
    \left(
    \frac{1}{x}
    \frac{d}{dx}
    x
    \frac{I_0(x)}
    {I_1(x)}
    \right)^{-1}
    \bigg|_{x=uy_r}
    -
    \frac{1}{16}
    u^2
    ,
    \end{align}
and using~\eqref{eq:26} express it as a function of the parameter $u$. The plot of the ratio ${\Delta\beta_g}/{\Delta\beta_{g0}}$ as a function of the parameter $a\omega_0/c\gamma$ is shown in Fig.~\ref{fig:3}.
%
\section{Calculation of the loss factor}\label{app:2}
%

In this Appendix we calculate the excitation of the resonant mode by a relativistic charge moving in a pipe with corrugated walls assuming $\gamma\gg 1$ but keeping small terms on the order of $\gamma^{-2}$, and using the boundary condition~\eqref{eq:16}.

Electric and magnetic fields of a point charge moving along the $z$ axis can be described with the electric potential $\varphi$ and the $z$-component $A_z$ of the vector potential $\vec{A}$. In the Lorentz gauge, $\p{\varphi}/{\p ct}+\p{A_z}/{\p z} = 0$, they satisfy the wave equations:
\begin{align}
  \label{eq:33}
  \nabla^2\varphi - \frac{1}{c^2}\parder{^2\varphi}{t^2}
  &=
  -4\pi q\, \delta(z-vt)\,\delta(\pi r^2)
  ,\nonumber \\
  \nabla^2A_z - \frac{1}{c^2}\parder{^2A_z}{t^2}
  &=
  -4\pi q\, (v/c)\,\delta(z-vt)\,\delta(\pi r^2)
  ,
\end{align}
where $r$ is the distance from the axis. We make the Fourier transformation in $z$ and time
    \begin{align}\label{eq:34}
    \hat \varphi(r,k_z,\omega)
    &=
    \int_{-\infty}^\infty
    dt\,
    dz\,
    e^{-ik_zz+i\omega t}
    \varphi(r,z,t)
    ,
    \nonumber\\
    \hat A_z(r,k_z,\omega)
    &=
    \int_{-\infty}^\infty
    dt\,
    dz\,
    e^{-ik_zz+i\omega t}
    A_z(r,z,t)
    .
    \end{align}
This transforms equations~\eqref{eq:33} into
  \begin{align}
  \label{eq:35}
  \frac{1}{r}\frac{d}{dr}r\frac{d}{dr}
  \hat \varphi
  +\left(\frac{\omega^2}{c^2}-k_z^2\right)
  \hat \varphi
  &=
  -8\pi^2 q\,\delta(\omega-k_zv)\,\delta(\pi r^2)
  ,\nonumber\\
  \frac{1}{r}\frac{d}{dr}r\frac{d}{dr}
  \hat A_z
  +\left(\frac{\omega^2}{c^2}-k_z^2\right)
  \hat A_z
  &=
  -8\pi^2 q\,(v/c)\,\delta(\omega-k_zv)\,\delta(\pi r^2)
  .
  \end{align}
A partial solution of these equations corresponding to the field in free space is $\hat\varphi = 4\pi q\delta(\omega-k_zv)K_0(|k_z|r/\gamma)$ and $\hat A_z = (v/c)\hat\varphi$, with $K_0$ the modified Bessel function of the second kind. To this partial solution we now add a general solution of the homogeneous equations   bounded at $r\to 0$:
\begin{equation}
  \label{eq:36}
  \hat\varphi 
  = 
  4\pi q\delta(\omega-k_zv)
  \left[K_0(|k_z|r/\gamma)
  +
  \alpha I_0(|k_z| r/\gamma)
  \right]
  ,
  \qquad
  \hat A_z = (v/c)\hat\varphi
  ,
\end{equation}
where $\alpha$ will be found from the boundary condition. 

The electric and magnetic fields involved into the boundary condition~\eqref{eq:16} are
    \begin{align}\label{eq:37}
    \hat E_z
    &=
    -ik_z \hat\varphi
    +
    \frac{i\omega}{c}
    \hat A_z
    =
    -\frac{ik_z}{\gamma^2}
    \hat\varphi
    \nonumber\\
    &=
    -\frac{4\pi q ik_z}{\gamma^2}
    \delta(\omega-k_zv)
    \left[K_0(|k_z|r/\gamma)
    +
    \alpha I_0(|k_z| r/\gamma)
    \right]
    ,
    \end{align}
and
    \begin{align}\label{eq:38}
    \hat H_\theta
    &=
    -
    \frac{\p \hat A_z}{\p r}
    =
    -
    \frac{4\pi |k_z|qv}{c\gamma}
    \delta(\omega-k_zv)
    \left[K_0'(|k_z|r/\gamma)
    +
    \alpha I_0'(|k_z| r/\gamma)
    \right]
    .
    \end{align}
Substituting these equations into~\eqref{eq:16} and using the expressions for the derivatives
    \begin{align}\label{eq:39}
    K'_0(x)
    =
    -
    K_1(x)
    ,\qquad
    I'_0(x)
    =
    I_1(x)
    ,
    \end{align}
we obtain
    \begin{align}\label{eq:40}
    \alpha
    &=
    \frac{
    \zeta(\omega)
    \beta
    K_1(|k_z|a/\gamma) 
    \mathrm{sign}(k_z)
    -
    i\gamma^{-1}
    K_0(|k_z|a/\gamma) 
    }
    {
    i\gamma^{-1}
    I_0(|k_z|a/\gamma)
    +
    \zeta(\omega)
    \beta
    I_1(|k_z|a/\gamma)
    \mathrm{sign}(k_z)
    }
    .
    \end{align}
In what follows we again  will use the approximation~\eqref{eq:21} for $\zeta$.

We now substitute $\alpha$ into~\eqref{eq:37}, select only the second term proportional to $\alpha$ (the first term is singular on the axis and describes the vacuum electric field of the moving charge), set $r=0$ and $z=vt$ and make the inverse Fourier transformation. This gives the longitudinal electric field acting on the particle, $E_{z0}=E_{z}(z=vt,r=0,t)$. Using the notations $u={a\omega_0}/{c\gamma}$ and $y=\omega/\omega_0$, and replacing $v\approx c$, we obtain 
    \begin{align}\label{eq:41}
    E_{z0}
    &=
    -
    \frac{iq}{\pi c^2\gamma^{2}}\,
    \int_{-\infty}^\infty
    \omega
    \alpha 
    \, d\omega
    =
    -
    \frac{iq}{\pi a^2}
    u^2
    \int_{-\infty}^\infty
    y
    \frac{
    2|y|K_1(u|y|) 
    +
    u
    K_0(u|y|) 
    }
    {
    -
    u
    I_0(u|y|)
    +
    2|y| I_1(u|y|) 
    }
    dy
    .
    \end{align}

\begin{figure}[htb]
\centering
\includegraphics[width=0.6\textwidth, trim=0mm 0mm 0mm 0mm, clip]{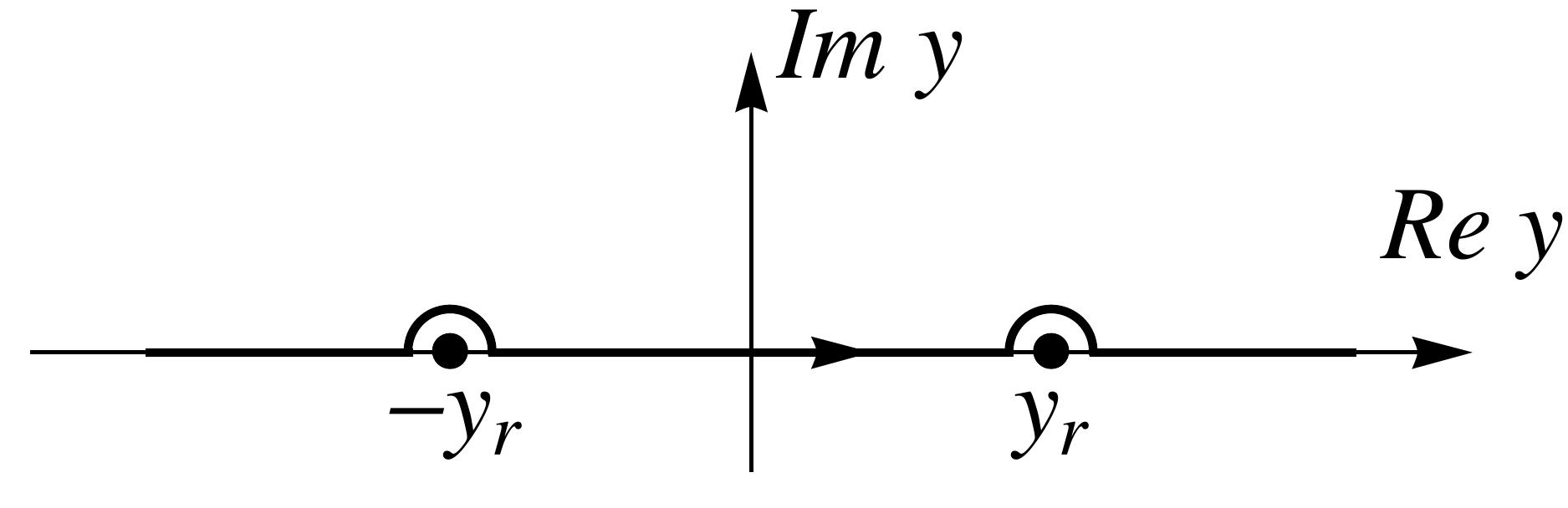}
\caption{Complex plane of variable $y$. Shown by the solid black line is the integration path in~\eqref{eq:41}.}
\label{fig:6}
\end{figure}

The integrand in~\eqref{eq:41} has poles on the real axis $y$ when its denominator vanishes. As one can see, the poles are located at $y=\pm y_r$ with $y_r$, determined  by Eq.~\eqref{eq:26}, that is by the condition that the phase velocity of the mode is equal to the velocity of the particle. These poles should be bypassed in the complex plane $y$ in accordance with rule that is established in the theory of the Cherenkov radiation~\cite{landau_lifshitz_ecm}. The rule can be easily understood if one introduces small losses into the boundary condition~\eqref{eq:16} by adding an infinitesimally small positive real part $\epsilon>0$ to $\zeta$, $\zeta \to \zeta + \epsilon$.  With account of $\epsilon$ the poles are shifted into the lower half plane of the complex variable $y$ and the integration path takes the shape shown in Fig.~\ref{fig:6}.  
The integral reduces to the sum of the half-residues from the poles (with the negative sign), and is given by the following expression:
    \begin{align}\label{eq:42}
    E_{z0}
    &=
    -
    \frac{2q}{ a^2}
    K
    \left(
    \frac{ak_0}{\gamma}
    \right)
    ,
    \end{align}
where the factor $K$ is    
    \begin{align}\label{eq:43}
    K(u)
    =
    u^2
    y
    \frac{
    2
    y
    K_1(uy) 
    +
    u
    K_0(uy) 
    }
    {
    d[
    -
    u
    I_0(uy)
    +
    2y
    I_1(uy)
    ]
    /dy
    }
    \bigg|_{y=y_r}
    .
    \end{align}
The loss factor is related to $E_{z0}$ through the equation $\varkappa=-E_{z0}/q$.
It is easy to see that in the limit $u\to 0$ the factor $K\to 1$ and we reproduce the result~\eqref{eq:4} for the loss factor in the limit $\gamma\to\infty$. The function $K(u)$ is plotted in Fig.~\ref{fig:4}.

%
\section{Derivation of the dispersion relation for the FEL instability of resonant mode in corrugated pipe}\label{app:3}
%

Starting with the Vlasov equation~\eqref{eq:5} it is convenient to introduce new variables: $\bar{s} =  k_w s$, where $k_w$ is defined by~\eqref{eq:6} and $\theta = \omega_r z/c$, and consider $f$ as a function of $\bar{s}$ and $\theta$. We linearize Eq. (\ref{eq:5}) assuming $f=f_0(\eta) + f_1(\eta, \theta, \bar{s})$ with $|f_1| \ll f_0$. Using notation $f_0(\eta) = n_0 F(\eta)$, where $n_0$ is the number of particles per unit length of the beam, we find
    \begin{align}\label{eq:44}
    &
    \frac{\partial f_1}{\partial \bar s}
    -
    \alpha\eta 
    \frac{\omega_r}{ck_w}
    \frac{\partial f_1}{\partial \theta}
    -
    (2\rho)^3
    F'(\eta)
    \int_{\theta}^{\theta+\bar s} d\theta'
    \int_{-\infty}^\infty d\eta'
    \nonumber\\
    &\times
    \tilde w(\theta'-\theta)
    f_1
    (\eta',\theta',\bar s-\theta'+\theta)
    =
    0
    ,
    \end{align}
where $\rho$ is the Pierce parameter \cite{bonifacio84pn} given by
    \begin{align}\label{eq:45}
    (2\rho)^3
    = 
    \frac{2n_0\kappa c r_0}{k_w\gamma\omega_r}
    ,
    \end{align}
and $\tilde w$ is the dimensionless wake expressed as a function of the dimensionless argument $\theta$,
    \begin{align}\label{eq:46}
    \tilde w(\theta)
    =
    \cos(\theta)
    .
    \end{align}

We then introduce a new variable ${\bar s}' = \bar{s}+\theta - \theta'$, and rewrite Eq. (\ref{eq:44}) in the following form
    \begin{align}\label{eq:47}
    &
    \frac{\partial f_1}{\partial \bar s}
    -
    \alpha\eta 
    \frac{\omega_r}{ck_w}
    \frac{\partial f_1}{\partial \theta}
    -
    (2\rho)^3
    F'(\eta)
    \int_{0}^{\bar s} d{\bar s}'
    \int_{-\infty}^\infty d\eta'
    \nonumber\\
    &\times
    \tilde w(\bar{s} - \bar{s}')
    f_1
    (\eta',\theta+\bar{s}-{\bar s}',{\bar s}')
    =
    0
    \,.
    \end{align}
We assume  a sinusoidal modulation of the distribution function with the wavenumber $k$, $f_1 \propto e^{i k z} = e^{i (1+ \nu) \theta}$, where $\nu = (ck - \omega_r)/\omega_r$ with $|\nu|\ll 1$. We then define functions $\Phi_\nu$ and $K_\nu$ such that
    \begin{eqnarray*}
    f_1(\eta, \theta, {\bar s})
    &=&
    e^{i(1+\nu)\theta}
    \Phi_\nu(\eta,{\bar s})
    \,,
    \nonumber\\
    K_\nu({\bar s})
    &=&
    e^{-i(1+\nu){\bar s}}\tilde w (\bar s)
    \,.
    \end{eqnarray*}
Eq. (\ref{eq:47}) takes the form
    \begin{align}\label{eq:48}
    \frac{\partial \Phi_\nu}{\partial \bar{s}}
    &
    -
    \alpha\eta 
    \frac{\omega_r}{ck_w}
    i (1+\nu)
    \Phi_\nu
    =
    (2\rho)^3
    F'(\eta)
    \int_0^{\bar s}
    d \bar{s}'
    K_\nu(\bar{s}' - \bar{s})
    \nonumber\\
    &\times
    \int_{-\infty}^\infty 
    d\eta'
    \Phi_\nu(\eta',\bar{s}' )
    =
    0
    \,.
    \end{align}
Laplace transforming Eq. (\ref{eq:48}) we find
    \begin{align}\label{eq:49}
    -&\Phi_\nu(\eta,0)
    -
    \alpha\eta 
    \frac{\omega_r}{ck_w}
    i (1+\nu)
    \tilde{\Phi}_\nu(\eta,q)
    =
    (2\rho)^3
    F'(\eta)
    \tilde{K}_\nu(q)
    \int_{-\infty}^\infty 
    d\eta'
    \tilde{\Phi}_\nu(\eta',q )
    \,,
    \end{align}
where
    \begin{align}\label{eq:50}
    \tilde{\Phi}_\nu(\eta,q)
    &=
    \int_0^\infty
    d{\bar s}
    e^{-q {\bar s}}
    \Phi_\nu(\eta,\bar{s})
    \,,
    \nonumber\\
    \tilde{K}_\nu(q)
    &=
    \int_0^\infty
    d{\bar s}
    e^{-q {\bar s}}
    K_\nu(-\bar{s})
    =
    \frac{1}{2}
    \left(
    \frac{1}{q  - i\nu}
    +
    \frac{1}{q  - i\nu - 2i}
    \right)
    \,.
    \end{align}
Dividing Eq. (\ref{eq:49}) by $q - i\alpha\eta (\omega_r/ck_w)(1+\nu)$ and integrating over $\eta$ yields
    \begin{align}\label{eq:51}
    \int_{-\infty}^\infty d\eta
    \tilde{\Phi}_\nu(\eta,q )
    =
    \frac{
    \int_{-\infty}^\infty 
    d\eta
    \frac{\Phi_\nu(\eta,0)}
    {q - i\alpha\eta(\omega_r/ck_w)(1+\nu)}
    }
    {
    1-(2\rho)^3
    \tilde{K}_\nu(q)
    \int_{-\infty}^\infty 
    d\eta
    \frac{F'(\eta)}{q - i\alpha\eta(\omega_r/ck_w)(1+\nu)}
    }
    \,.
    \end{align}
The dispersion relation that defines the propagating constant $q$ of the mode is given by zeros of the denominator on the right hand side of this equation:
    \begin{align}\label{eq:52}
    (2\rho)^3
    \tilde{K}_\nu(q)
    \int_{-\infty}^\infty d\eta
    \frac{F'(\eta)}{q - i\alpha\eta (\omega_r/ck_w)(1+\nu)}
    =
    1\,.
    \end{align}
Rapid growth will be seen to correspond to $|\nu| \lesssim 2\rho$ and $q \sim 2\rho$. The second term in expression for $\tilde{K}_\nu$ in~(\ref{eq:50}) is not resonant and can be neglected, which gives
    \begin{align}\label{eq:53}
    \frac{1}{2}
    (2\rho)^3
    \frac{1}{q - i\nu}
    \int_{-\infty}^\infty 
    d\eta
    \frac{F'(\eta)}{q - i\alpha\eta({\omega_r}/{ck_w})}
    =
    1
    \,,
    \end{align}
where we neglected $\nu$ relative to unity in the denominator of the integrand of Eq. (\ref{eq:52}).

\end{document}